\documentclass{article}

\usepackage{arxiv}

\usepackage[utf8]{inputenc} 
\usepackage[T1]{fontenc}    
\usepackage{hyperref}       
\usepackage{url}            
\usepackage{booktabs}       
\usepackage{amsfonts}       
\usepackage{nicefrac}       
\usepackage{microtype}      
\usepackage{lipsum}		
\usepackage{graphicx}
\usepackage{natbib}
\usepackage{doi}

\usepackage{comment}
\usepackage{todonotes}
\usepackage{graphicx}

\usepackage{amsmath}
\usepackage{algorithmic}
\usepackage{array}

\usepackage{stfloats}

\usepackage{xcolor}
\usepackage{upgreek}

\usepackage{rotating}
\usepackage{amsmath}
\usepackage{multirow}
\usepackage{graphicx}
\graphicspath{{figures/}}
\usepackage[compress]{cite}
\usepackage{caption}
\usepackage{subcaption}
\usepackage{pifont}
\newcommand{\cmark}{\ding{51}}%
\newcommand{\xmark}{\ding{55}}%

\usepackage{pdflscape}
\usepackage{rotating}
\normalsize

\usepackage{booktabs}   

\usepackage{tabularx}
\usepackage{ltablex}
\usepackage{caption}
\usepackage{array}

\usepackage[utf8]{inputenc}
\usepackage[T1]{fontenc}
\usepackage{ltablex}
\usepackage{caption, booktabs}
\usepackage{newtxtext,newtxmath}

\usepackage{amsmath}
\usepackage{mathdots}
\usepackage{cancel}
\usepackage{color}
\usepackage{array}
\usepackage{multirow}
\usepackage{gensymb}
\usepackage{tabularx}
\usepackage{extarrows}
\usepackage{booktabs}
\usetikzlibrary{fadings}
\usetikzlibrary{patterns}
\usetikzlibrary{shadows.blur}
\usetikzlibrary{shapes}

\title{Thermal Heating in ReRAM Crossbar Arrays: Challenges and Solutions}

\author{ \href{https://orcid.org/0000-0001-6932-188X}{\includegraphics[scale=0.06]{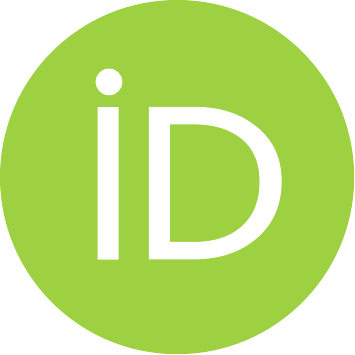}\hspace{1mm}Kamilya Smagulova}\\
	Division of CEMSE,\\
	KAUST\\
	Thuwal, KSA \\
	\texttt{kamilya.smagulova@kaust.edu.sa} \\
	\And
	\href{https://orcid.org/0000-0001-7139-3428}{\includegraphics[scale=0.06]{orcid.pdf}\hspace{1mm}Mohammed E. Fouda} \\
	Rain Neuromorphics, Inc.  \\
        San Francisco, CA, USA\\
	\texttt{foudam@uci.edu} \\
         \And
	\href{https://orcid.org/0000-0003-1849-083X}{\includegraphics[scale=0.06]{orcid.pdf}\hspace{1mm}Ahmed Eltawil} \\
	Division of CEMSE,\\
	KAUST\\
	Thuwal, KSA \\
	\texttt{ahmed.eltawil@kaust.edu.sa} \\
}

\renewcommand{\shorttitle}{\textit{arXiv} Template}
\hypersetup{
pdftitle={A template for the arxiv style},
pdfsubject={q-bio.NC, q-bio.QM},
pdfauthor={David S.~Hippocampus, Elias D.~Striatum},
pdfkeywords={First keyword, Second keyword, More},
}

\begin{document}
\maketitle

\maketitle

\begin{abstract}
The higher speed, scalability and parallelism offered by ReRAM crossbar arrays foster development of ReRAM-based next generation AI accelerators. At the same time, sensitivity of ReRAM to temperature variations decreases R$_{ON}$/R$_{OFF}$ ratio and negatively affects the achieved accuracy and reliability of the hardware. Various works on temperature-aware optimization and remapping in ReRAM crossbar arrays reported up to 58\% improvement in accuracy and 2.39$\times$ ReRAM lifetime enhancement. This paper classifies the challenges caused by thermal heat, starting from constraints in ReRAM cells' dimensions and characteristics to their placement in the architecture.  In addition, it reviews available solutions designed to mitigate the impact of these challenges, including emerging temperature-resilient DNN training methods. Our work also provides a summary of the techniques and their advantages and limitations.

\end{abstract}


%

\section{Introduction}
\label{sec:intro}

The rapid progress in 
artificial intelligence (AI) is dictating new requirements for hardware accelerators. Modern computational processes are characterized by an abundance of dot-product operation and an extreme lack of storage space. In this regard, non-volatility, nanoscale size and the ability to retain multiple states made \textit{resistive switching materials (RSMs)} promising in the design of energy-efficient high-density memory devices. Moreover, RSMs, and resistance random access memory (ReRAM) in particular, can act as synapses and allow building of artificial neurons and even neural networks. Multiple ReRAM cells organized into crossbar arrays can perform vector-matrix multiplication (VMM) faster and more efficiently than von-Neumann-based architecture since ReRAM cells can store data values as conductance states and reduce data movement 
between separate memory and processing units \citet{hu2018memristor}.  
Therefore, computing-in-memory (CIM) or processing-in-memory (PIM) analog and digital ReRAM-based accelerators such as ISAAC \citet{shafiee2016isaac}, PRIME \citet{chi2016prime}, PUMA \citet{ankit2019puma} and others are firmly entering modern electronics. In particular, ISAAC outperformed its fully digital counterpart DaDianNao with improvements of 14.8$\times$ in throughput, 5.5$\times$ in energy and 7.5$\times$ in computational density \citet{shafiee2016isaac}. 
\par 
\begin{figure}[hh]
\centering
\begin{minipage}[t]{.45\textwidth}
    \centering
    \includegraphics[width=1\textwidth]{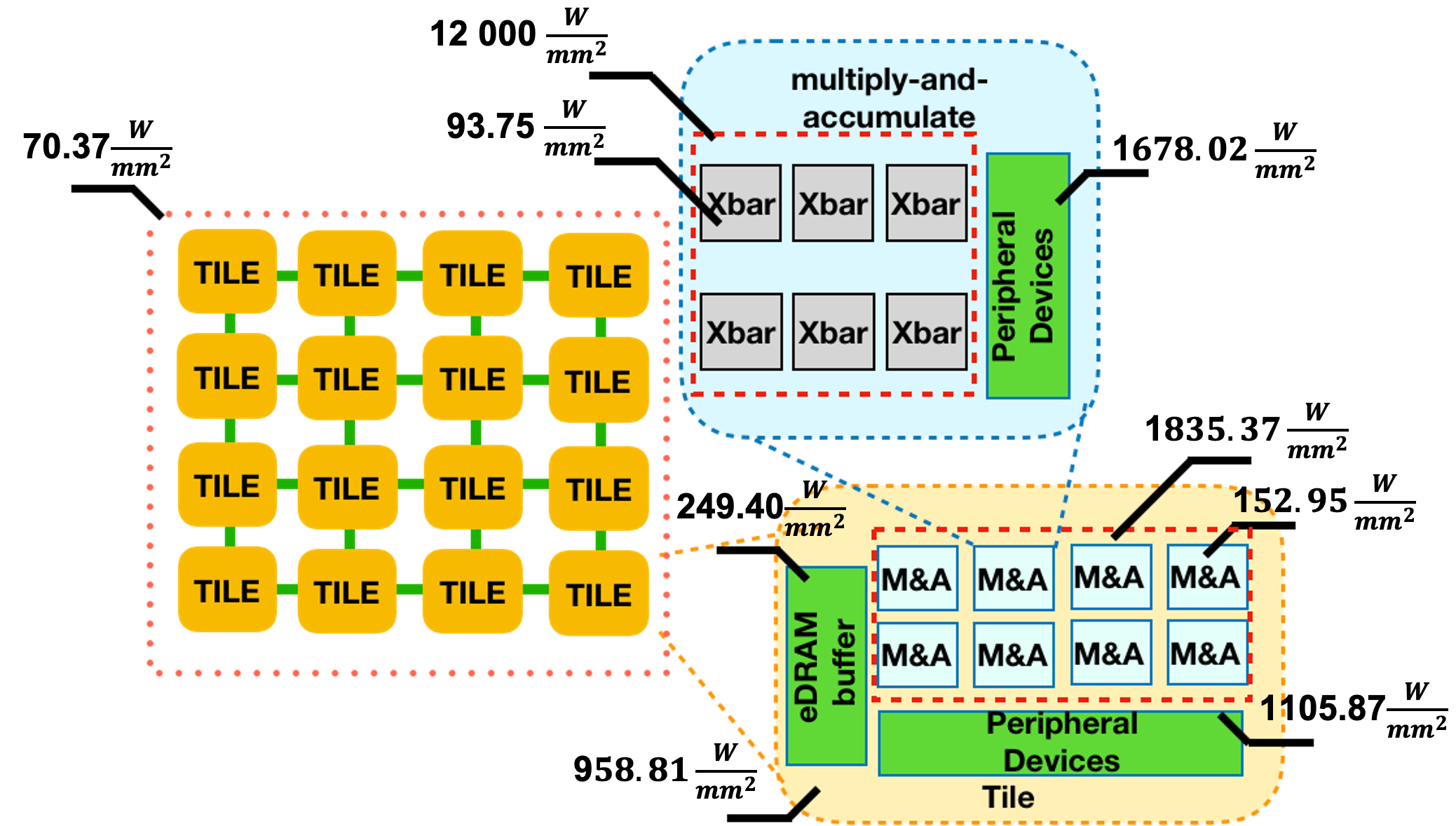}
    \subcaption{}\label{fig:1a}
\end{minipage}
\vspace{0.5cm}
\begin{minipage}[t]{.5\textwidth}
    \centering
    \includegraphics[width=1\textwidth]{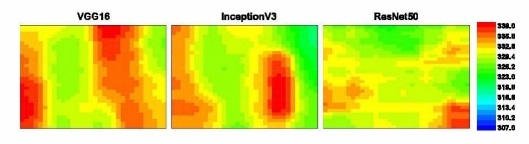}
    \subcaption{}\label{fig:2b}
\end{minipage}
\caption{ a) Power density of ISAAC-CE\citet{smagulova2021resistive};
b) Steady-state temperature distributions of the same ReRAM chip running three different CNN models for inference of ImageNet: VGG16, InceptionV3, ResNet50 \citet{liu2019hr}. }
\label{conduction1}
\vspace{-0.1in} 
\end{figure}

Nevertheless, an intra-class comparison with state-of-the-art (SoTA) commercial accelerators shows that existing ReRAM-based accelerators have higher power density with non-uniform distribution \citet{smagulova2021resistive}. On the other hand, it also leads to disproportional temperature distribution.  Previous works showed that an increase of temperature has an impact on resistive switching behavior and the R$_{ON}$/R$_{OFF}$ ratio of ReRAM cells \citet{walczyk2011impact}, \citet{beigi2019thermal}. In turn, change of the conductance states affects the accuracy of a ReRAM-based hardware \citet{beigi2018thermal}. Moreover, the materials and dimensions of ReRAM cells can define the level of the hardware's sensitivity to temperature \citet{sun2015thermal}, \citet{al2020reliability}. 
A closer look at ReRAM-based architectures shows that heterogeneous parts of the accelerators demonstrate non-uniform power density distribution (as shown in Figure \ref{conduction1}a for ISAAC)
and consequently result in uneven temperature regions  \citet{smagulova2021resistive}, \citet{zhang2022thermal}. Since power-hungry components have a higher rate of heat dissipation, they might interfere with the temperature and performance of surrounding elements. It was observed that an increase of temperature from 300K
to 400K
may reduce the accuracy of ReRAM-based hardware by up to 6$\times$ \citet{beigi2018thermal}. Moreover, due to different conductance values and input voltages, there might be a non-uniform temperature distribution within the ReRAM crossbar arrays too. Figure \ref{conduction1}b shows the steady-state thermal distribution in the same ReRAM chip during inference of VGG16, InceptionV3 and ResNet50 workloads for ImageNet dataset classification. As can be seen, the temperature difference between the models can reach up to 17.16K \citet{liu2019hr}. 
\par 
However, the majority of resistive hardware accelerators did not consider the thermal sensitivity of ReRAM cells in their design. The study of temperature impact on ReRAM-based architectures and the development of solutions to mitigate the problem started gaining attention only recently \citet{beigi2019thermal}, \citet{sun2015thermal}. 
This paper contributes in the following ways:
\begin{itemize}
\item we summarized the design and performance challenges of ReRAM-based hardware caused by temperature increase;
\item we reviewed existing solutions developed to address the identified challenges;  
\item  we categorized methods designed to mitigate the impact of temperature and analyzed their advantages and shortcomings compared to each other;
\item finally, based on the solutions discussion, we highlighted the key takeaways. 
\end{itemize}

The rest of the paper is organized as follows: Section \ref{sec:reram_accel} provides information on ReRAM crossbar arrays and SoTA ReRAM-based neural accelerators. Section \ref{sec:challenges} discusses the thermal challenges in ReRAM-based hardware caused by temperature increase and Section \ref{sec:solutions} introduces existing techniques developed to address these challenges. Finally, Section \ref{sec:discussion} provides a summary discussion of the presented solutions.

\par 






\section{Existing Resistive Neural Accelerators }
\label{sec:reram_accel}

\subsection{ReRAM Crossbar Arrays}

ReRAM is a non-volatile memory device with conducting filament (CF) material sandwiched between top and bottom electrodes \citet{chen2020reram}.  \textit{Resistivity swtiching} (RS) of ReRAM cells from High Resistance State (HRS) to Low Resistance State (LRS) and vice versa can be controlled via connection and disconnection of the CF. 
Typically ReRAM devices operate in \textit{read} and \textit{write} modes. During write mode, current or voltage pulses of certain amplitude, polarity and duration are applied to the ReRAM to program its state. Sensing the ReRAM state is performed during read mode via applying voltages and currents of a specific range.   
\begin{figure}[hh]
\centering
    \begin{subfigure}[]{0.23\textwidth}
        \includegraphics[width=1\textwidth]{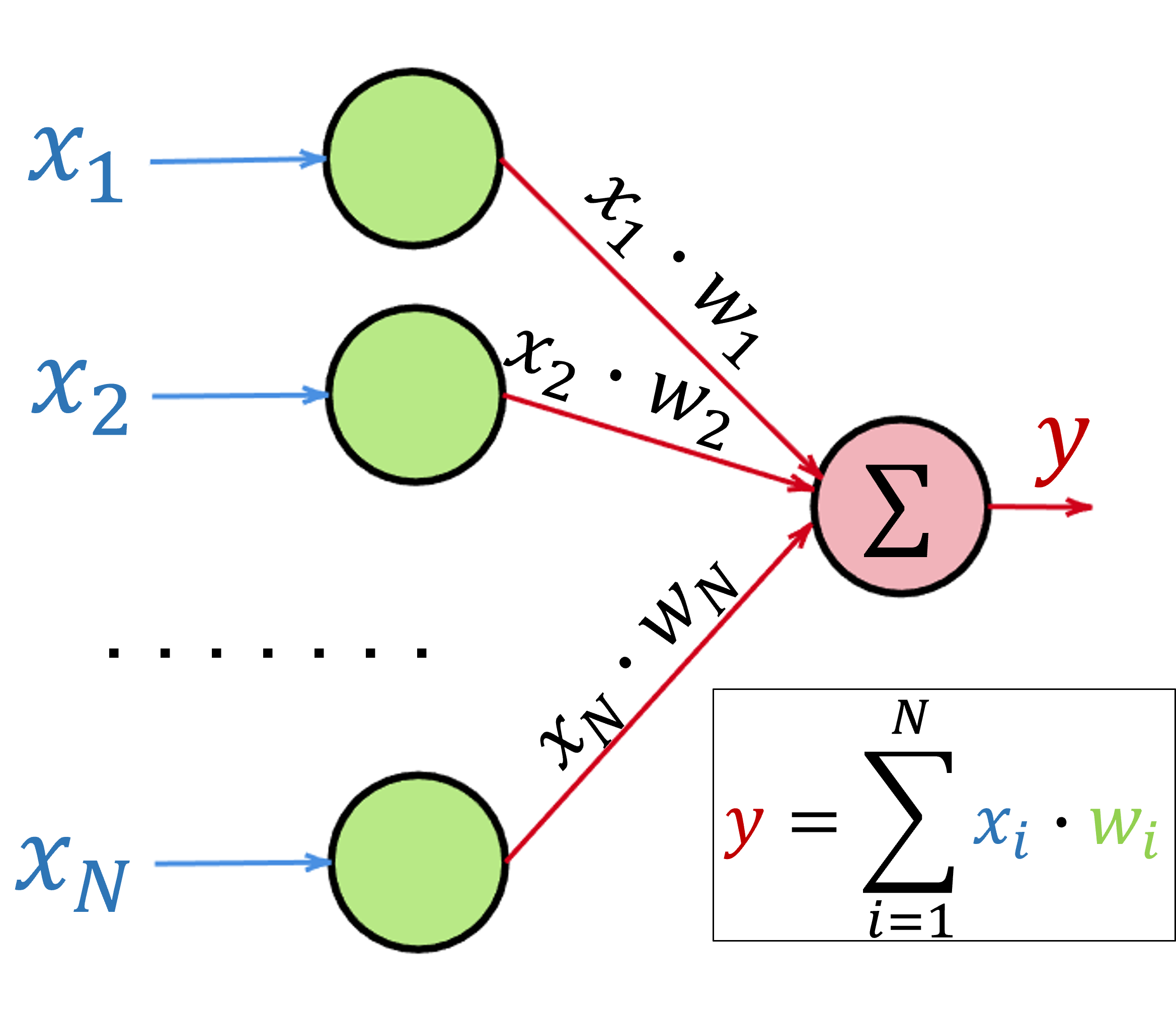}
        \caption{}
    \end{subfigure}
    \begin{subfigure}[]{0.32\textwidth}
        \includegraphics[width=1\textwidth]{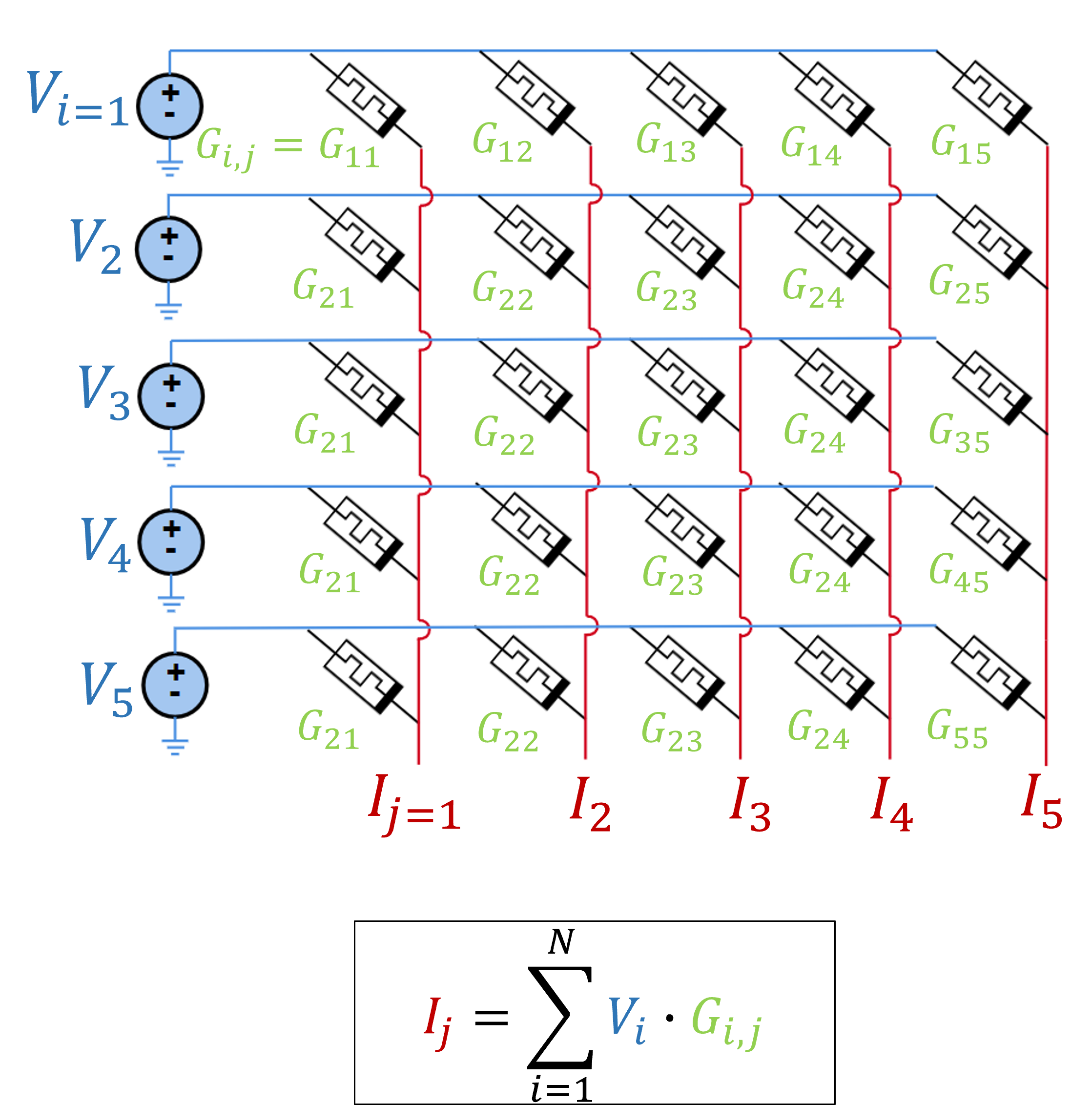}
        \caption{}
    \end{subfigure}

\caption{a) An artificial neuron; b) a ReRAM crossbar array.}
\label{crossbar}
\end{figure}

ReRAM's high speed and scalability, power efficiency, nanoscale size and ability to retain a value in a non-volatile manner sparked interest in ReRAM-based resistive crossbar array (RCA) architectures. RCAs can serve either as non-volatile memory devices for storing data or as CIM architectures for performing VMM or accelerating neural networks. In the latter application, a ReRAM cell acts as a synaptic weight $w_{i,j}$ of a neural network with a neuron output $y_{j} = \sum_{i=1}^{N} w_{i,j}\times x_{i}$ as shown in Figure \ref{crossbar}a. According to Kirchoff's current law (KCL), the output current of each column in RCA is equal to a weighted summation of the input voltages $I_{j} = \sum_{i=1}^{N} G_{i,j}\times V_{i}$ (Figure \ref{crossbar}b). This property forms the basis of many ReRAM-based accelerators \citet{hu2018memristor}.

\subsection{SoTA ReRAM Accelerators}

\begin{figure*}[ht]
\centering
   \includegraphics[width=0.95\textwidth]{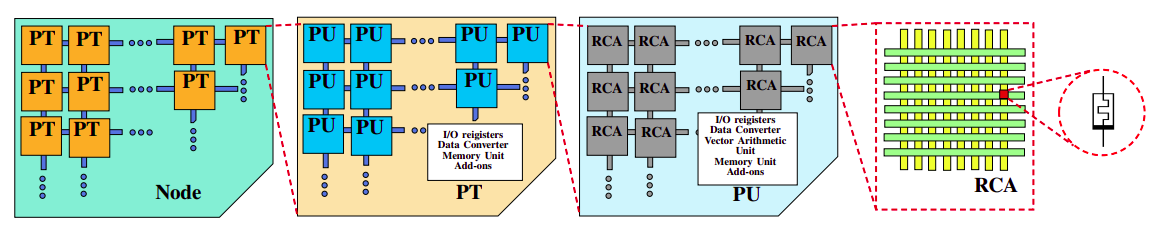}
\caption{Hierarchical architecture of a ReRAM-based CIM accelerator (2D planar design).} 
\label{hierar}
\end{figure*}

The typical architecture of many-core bank- or tile-based resistive hardware accelerators comprises ReRAM crossbar arrays and various peripheral circuits 
and interconnects. Two of the first many-core ReRAM-based accelerator designs were ISAAC \citet{shafiee2016isaac} and PRIME \citet{chi2016prime}. ISAAC has many-core architecture with tiles connected via network-on-chip (NoC). Compared to fully digital neural network accelerator DaDianNao \citet{chen2014dadiannao}, utilization of RCAs in ISAAC for VMM operation allowed ISAAC to reduce energy by 5.5$\times$ and increase throughput and computational density by up to 14.8$\times$ and 7.5$\times$, respectively  \citet{shafiee2016isaac}.  PRIME consists of banks that are connected via bus interconnect and uses RCAs for both data storage and VMM. Both accelerators support only the inference phase with 16-bit precision. Figure \ref{hierar} shows the hierarchical structure of a ReRAM-based accelerator, including a node, processing tile (PT), processing unit (PU) and RCA. Communication between on-chip and off-chip components in multi/many-core platforms takes place via interconnects. Currently, ISAAC and PRIME serve as baseline models for the majority of SoTA ReRAM neural accelerators. Subsequent architectures such as AEPE \citet{tang2017aepe}, PUMA \citet{ankit2019puma}, Newton \citet{nag2018newton} and others have mainly aimed to decrease the power consumption of the peripheral circuits by modification of the resolution of ADC and DAC circuits or optimization of the weight mapping. In addition, PRIME-based PipeLayer \citet{song2017pipelayer} and ISAAC-based AtomLayer \citet{qiao2018atomlayer} and PANTHER \citet{ankit2020panther} architectures provide support for on-chip training phases.
\par 

\begin{figure*}[ht]
\centering
    \begin{minipage}[b]{0.25\textwidth}
         \centering 
        \subfloat[]
        {\includegraphics[width=0.55\textwidth]{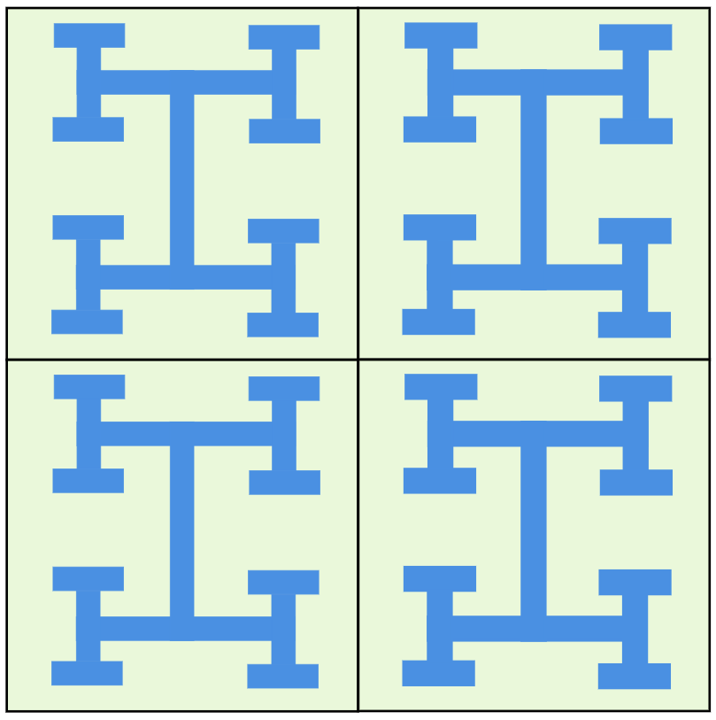}
        \label{fig:CNN}}
        \vspace{8mm}
        \subfloat[]
        {\includegraphics[width=1\textwidth]{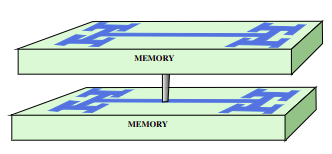}
        \label{fig:TSV_3D}}
    \end{minipage}
  \hspace{5mm}
    \begin{minipage}[b]{0.40\linewidth}
        \centering        
        \subfloat[]
        {\includegraphics[width=1\textwidth]{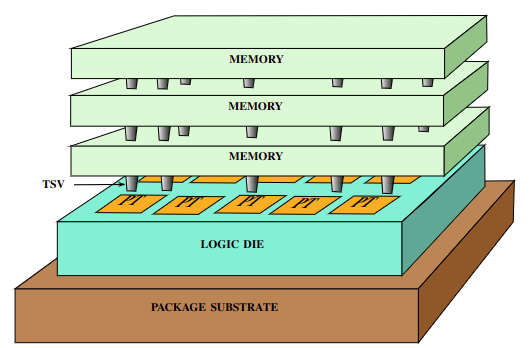}}
        \label{fig:HMC}

    \end{minipage}  
    \hspace{8mm}
      \begin{minipage}[b]{0.25\textwidth}
        \centering 
        \subfloat[]
        \centering {\includegraphics[width=0.95\textwidth]{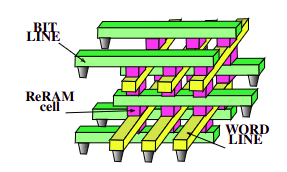}
        \label{fig:HRRAM}}
        \subfloat[] {\includegraphics[width=0.9\textwidth]{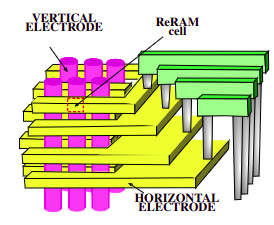}
        \label{fig:VRRAM}}
    \end{minipage}
\caption{a) 2D memory architecture with H-tree routing path (top view); b) TSV-3D memory architecture (side view); c) TSV-based Hybrid Memory Cube (HMC) ReRAM accelerator: Memory composed of eDRAM and Logic die composed of processing tiles (PTs);  d) H-ReRAM crossbar array:  e) V-ReRAM crossbar array.} 
\label{crossbar2}
\end{figure*}
\par
Heterogeneous on-chip and off-chip components in multi/many-core platforms should be placed to ensure high signal transmission speed/rate, small area and low power. Traditional two-dimensional integrated circuits (2D ICs) (Figure \ref{hierar}) are no longer feasible for this task and active research is being conducted in the fields of 2.5D/3D stacking (Figure \ref{crossbar2}b-e) \citet{lau2021semiconductor}. A through-silicon via (TSV) (also called an active TSV-interposer) technology allows bonding several dies in a face-to-face (F2F), face-to-back (F2B) and back-to-back (B2B) manner. TSV is used in 2.5D/3D die-stacking, including popular commercial technologies like Micron's Hybrid Memory Cube (HMC) and Hynix's High Bandwidth Memory (HBM). However, TSV does not scale well as the technology node size shrinks. Recently proposed monolithic three-dimensional (M3D) integration, also called 3D sequential integration, allows integration of ICs on top of each other on a single silicon substrate \citet{cheng2022emerging}. One of the metrics for comparing integration technology is PPC:
\begin{equation} 
\text{PPC}= \text{performance}/(\text{power}\times \text{cost}))
\end{equation}
In terms of PPC, TSV-based 3D can achieve only half of M3D gain \citet{cheng2022emerging}.  
Improvement in bandwidth and power can also be achieved by stacking 2D planar ReRAM crossbar arrays into horizontal 3D ReRAM (H-ReRAM) or horizontal cross-point architecture (HCPA). There is also a vertical 3D ReRAM (V-ReRAM) design known as a vertical cross-point architecture (VCPA). Here, multiple devices are fabricated at the sidewall of horizontally running wordlines (WL) and a vertically oriented bitline (BL). Both H-ReRAM and V-ReRAM  allow scaling the ReRAM device size down to 4$F^{2}/n$ where $n$ is the number of stacked layers \citet{hudec20163d}. Generally, 3D die-stacking and 3D stacking of ReRAM arrays also lead to higher power densities and thermal problems. 

\begin{figure}[hh]
\centering
\includegraphics[width=0.4\textwidth]{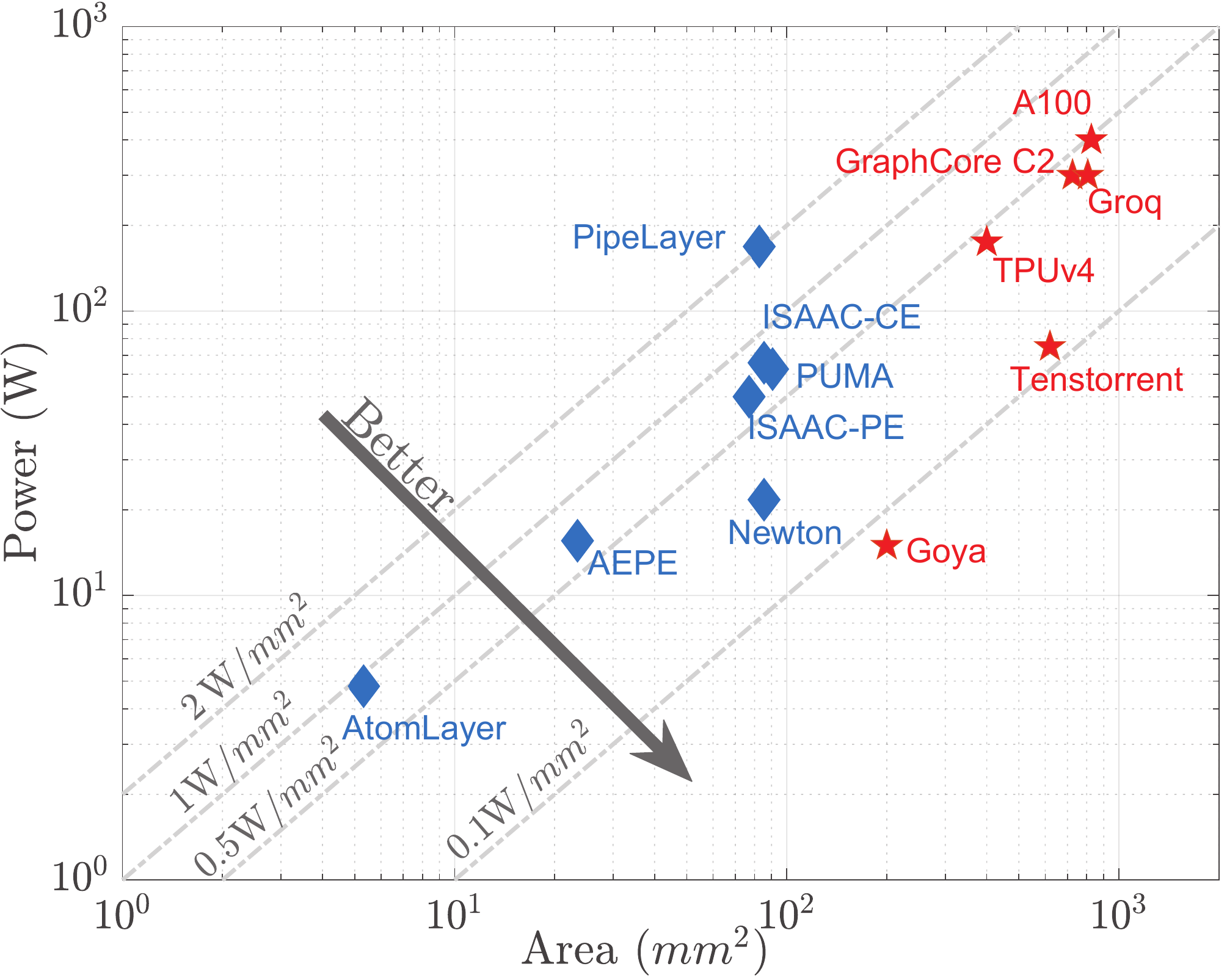} 
\caption{Power density of the state-of-the-art ReRAM-based and commercial  accelerators.}
\label{W_mm2}
\end{figure}
In Figure \ref{W_mm2}, SoTA ReRAM neural accelerators are compared against commercial accelerators, including  Goya\citet{medina2020habana}, Google TPUv4\citet{wang_selvan}, GraphCore C2\citet{lacey}, Groq\citet{gwennap2020groq}, Nvidia A100 \citet{campa_kawalek_vo_bessoudo_2021} and Tenstorrent \citet{Tenstorrent}. An intra-class comparison shows that the power density of the commercial accelerators is always less than $0.5W/mm^2$, whereas the power density of the majority of resistive accelerators is above the bound and reaches $2W/mm^2$ in the case of PipeLayer.

\section{Challenges}
\label{sec:challenges}

All processes that take place in ReRAM during RS are thermally activated and can be described using an Arrhenius dependence \citet{sun2015thermal}. Due to the accumulative effect at the output of RCAs, the ReRAM conductance state drift caused by temperature increase severely affects the performance of the ReRAM-based DNN accelerators rather than ReRAM-based storage devices.  The challenges associated with temperature variation and thermal disturbance in ReRAM-based hardware are provided below.

\par

\subsection{Challenge 1: Static and Dynamic Retention}

Study of the conduction mechanism in RSMs has shown that room temperature affects the readout margin of a device \citet{walczyk2011impact}. In particular, a stable bipolar switching behavior in TiN/HfO$_{2}$/Ti/TiN ReRAM on 0.25$\mu m$ complementary metal–oxide–semiconductor (CMOS) technology is observed within the temperature range 213–413K. 
However, further temperature increase leads to a proportional decrease of R$_{OFF}$/R$_{ON}$ ratio and data loss. To describe the temperature effect on ON-state and OFF-state, a quantum point-contact (QPC) framework was used. According to the QPC model, the ON-state shows a metallic characteristic and resistance can be modeled as: 
\begin{equation}
    R_{ON} = R^{0}_{ON} [ 1+ \rho (T - T_{0}) ]
\end{equation}
where  $R^{0}_{ON}$ is the resistance measured at temperature $ T_{0}$= 293K; and temperature coefficient $\rho$ = 3$\times$$10^{-2}$ $1/K$.  



Another test was conducted on 2-bit 256$\times$256 1T1R HfO$_{2}$-based array 90nm technology. The temperature was varied from 300K to 395K and the measured \textit{static retention} characteristics were used to update a model \citet{chen2015compact}, resulting in the following equations \citet{shim2021impact}:
\begin{align}
\Delta\mu = \mu(t)-\mu_{init} = A_{avg}\times\log t\\
\Delta\sigma = \sigma(t)-\sigma_{init} = B_{var}\times\log t
\end{align}
where $t$ is retention time; $\mu$ is the average conductance of the state and $\sigma$ is its standard deviation; and $A_{avg}$ and $B_{avg}$ are the conductance drift rates that depend on temperature. According to the observations, the intermediate states of ReRAM are more susceptible to thermal effect \citet{chen2015compact},\citet{wang2018conduction}. 
\par 
In addition to a static retention variation, there is a \textit{dynamic retention} variation caused by temporal temperature changes. Dynamic retention can be modeled as the sum of the static
variations at each temperature step \citet{meng2021temperature}.


\subsection{Challenge 2: Relaxation}

The short-term relaxation effect 
is a random conductance drift of a ReRAM back to the initial state that takes place right after the state programming and has a significant impact on the overall accuracy \citet{xi2020impact}. To investigate a \textit{volatile jump} in ReRAM, a change of resistance $R(t)$ within a given time window $t_{w}$ with $R_{start} = R(0)$ and $R_{end} =R(t_{w})$ was measured during write and read phases. Assuming that $R_{pre}$ is the initial resistance, a volatile change can be determined as:  
\begin{equation}
    \Delta R_{start} = R_{start} - R_{pre}
\end{equation}
and $R$ lasting after $t_{w}$ is defined as a \textit{non-volatile} residue:
\begin{equation}
    \Delta R_{end} = R_{end} - R_{pre}
\end{equation}
The resulting ReRAM volatility model over time \textit{t} can be expressed as follows:
\begin{equation}
R(t) = \alpha \exp ^{\left(-\frac{t}{\uptau}\right)^\beta} + \gamma
\end{equation}
where $\uptau$ is the relaxation time constant; $\alpha$ is the relative offset between the initial resistive state and the predicted non-volatile saturation point; $\beta$ is the stretch factor; and $\gamma$ is the non-volatile saturation point \citet{giotis2020bidirectional}. 

\par 
The majority of teh research is concentrated on the effect of temperature on the \textit{non-volatile} switching behavior of ReRAMs including long-term retention. Recent work has shown that temperature also affects the \textit{volatile} resistive state and the corresponding relaxation time constant $\uptau$\citet{abbey2022thermal}.  Three $TiO_{x}$-based ReRAMs were tested under temperatures T = [295, 313, 268, 343]K for a range of write pulses $V_{P} \in$ ± [1.5, 5.0]V. The magnitude of volatile jumps was proportional to the applied voltage. However, in the case of positive amplitude pulses, there was an increase in variability toward 5V. In the case of negative amplitude pulses, an increase in voltage led to saturation. Temperature $T$ rise also had an opposite effect depending on voltage polarity.  The inhibition effect took place at the positive polarity and low bias amplitude as well as in the case of positive polarity and high bias voltages. Otherwise, enhancement effects occurred\citet{xi2022impact}.


\subsection{Challenge 3: Endurance}

The \textit{expected shortest lifetime (ESL)} of ReRAM is around 8 years. From prior works \citet{strukov2016endurance} the dependence of endurance of temperature variation can be expressed via write latency $t_{w}$ \citet{beigi2019thermal} :
\begin{equation}
Endurance \approx \left(\frac{t_{w}}{t_{0}}\right)^{\frac{U_{F}}{U_{S}} - 1}
\end{equation}
where $t_{0}$ is a constant that depends on the device; and $U_{F}$ and $U_{S}$ are the activation energy for the failure mechanism and the switching mechanism, respectively. For non-volatile devices, the typical ratio of $\frac{U_{F}}{U_{S}}$ varies from 2 to 4. From the analytical model, it was derived that increasing the temperature from 300K
to 330K 
decreases  $t_{w}$ from 50 $ns$ to 30 $ns$ 
and reduces device endurance \citet{beigi2019thermal}. Moreover, high rates of SET-RESET also increase the temperature and decrease the average ReRAM lifetime. Surprisingly, low temperature also has a negative impact on ReRAM as it hinders recovery of a broken filament \citet{fadeev2021issue}.

\subsection{Challenge 4: Thermal Cross-Talk Effects}
The repeated SET-RESET switching cycles in a ReRAM device generate Joule heat, which may also affect the performance of surrounding devices. To quantify the thermal effect, a $Cu/TaO_{x}/Pt$ crossbar array with a neighboring line pitch of between 150 $\mu$m and 185 $\mu$m was studied. The performance of heated ("aggressor") and unheated ("victim") devices can deteriorate to a certain degradation factor $D$, which can be found from the following equation: 

\begin{equation}
  D = 1-\frac{N(unheated)}{M_{x}(heated)}  
\end{equation}
where $M_{x}$ is the maximum number of switching cycles of a 'marginal' device. 
Testing of around 100  devices  showed that the $M_{x}$ of a $TaO_{x}$-based "marginal" device is around 13 and afterward the device demonstrates unstable performance \citet{al2020reliability}.

\par
In two-dimensional (2D) crossbar arrays, the degradation factor depends on the presence of a shared electrode, its material and size, and the remoteness of the unheated device from a heated cell. The study also showed that the first neighbor cell suffers the highest degree of degradation.
In particular, in $Cu/TaO_{x}/Pt$ crossbar arrays, degradation of the first neighbor along the $Pt$ electrode was about D=67\%; along the $Cu$ electrode it was 80\%. In the case of non-shared electrodes, degradation of the first diagonal neighbor was D=19\%. Thermal effect increases with downscaling of the pitch size and spacing between them. It becomes a huge issue in commercial ReRAM devices that are 1000$\times$ smaller than the studied device\citet{al2020reliability}. 

   

Another parameter that can be used to evaluate thermal cross-talk in a crossbar array is the time $t_{s}$ required to reach a thermal steady state. For an individual device with feature size 80$nm$, $t_{s}$ is 5 $ns$, which is less than the required RESET time. But for 1D1R cell in 1$\times$1$\times$1 array, $t_{s}$ is around 50$ns$ and steady-state temperature is equal to 500K, whereas in a 3$\times$3$\times$3 block array, $t_{s}$ is 500$ns$ and the temperature is 605K. Therefore, the thermal model of a single device should be extended.  
In \citet{sun2015thermal} the authors presented two different "worst case" scenarios - one in a typical crossbar array structure and the other in a crossbar array with shared WL/BL. In the first case, when ReRAM cells were reset from LRS to HRS by applying a reset pulse, a thermal heat propagated along the vertical direction and disturbed the unprogrammed layer. In the second case, the configuration allows erasing/programming at different layers of the crossbar.  This time heat from neighboring cells propagated in both vertical and horizontal directions and disturbed unprogrammed cells.





\subsection{Challenge 5: Die-Stacking}

ReRAM crossbar arrays can be stacked into heterogeneous structures using 2.5D and 3D integration technologies. These include TSV-based interposer and monolithic integration. The common interfacing methods in TSV-based integration are HBM and HMC. Such multiple die-stacking offers numerous advantages over 2D geometry scaling, includinge shorter interconnect, reduced latency, higher density and smaller footprint \citet{yu2017energy, dhananjay2021monolithic}. 

\par 
However, due to the different thermal densities of the components, stacked architectures suffer from inter-die thermal coupling and hotspots. 
Consequently, die-stacking leads to accuracy degradation and reliability challenges, including retention, thermal cross-talk and endurance. For instance, in a 2.5D stacking design 
the temperature in the ReRAM banks reaches up to 
344K and   decreases their lifetime close to or below ESL. In a 3D interposer stacking design, the vertical heating  
temperature rises up to 380K
and reduces ReRAM lifetime below 2.6 years \citet{beigi2019thermal}. In terms of die-to-die interconnections, M3D design has less area overhead  compared to TSV-3D, but it is more sensitive to temperature. At 10 years, the accuracy drop in M3D-air architecture was 53\%, whereas in TSV-3D it was 10\%\citet{kaul2021thermal}.

\subsection{Challenge 6: Limited Scaling Potential}


ReRAM among other NVM technologies is known for having the smallest size, around 4$F^{2}$. Typically, a single ReRAM size is below 10$nm$. Although ReRAM miniaturization allows saving power and area, scaling down the feature size ($F$) in devices such as $NiO$ ReRAM from 100$nm$ to 30$nm$ node can lead to an increase of temperature from around 400K up to 1800K. 
 In addition, miniaturization enhances the thermal cross-talk issue \citet{sun2015thermal}.


The thermal reaction model from \citet{sato2007consideration} was utilized to study the behavior of saturated temperature in various ReRAM devices at low resistivity (10 $\mu\Omega$ cm), medium resistivity (50 $\mu\Omega$ cm) and high resistivity (100 $\mu\Omega$ cm)  \citet{dongale2016investigating}. In the analysis, the reset voltage was set at 0.5V and the thickness of the oxidation membrane was 200nm. The radius of the conductive filament of $ZnO, TiO_{2}, WO_{3}$ and $HfO_{2}$ was varied from 10$nm$ to 100$nm$. 
\par 
Overall, the conduction mechanism in ReRAM is mainly defined by the material and the geometry of CF and electrodes in a metal-insulator-metal (MIM) structure. Most popular ReRAMs can be classified into conductive bridge random access memory (CBRAM) and metal oxide ReRAM (OxRRAM).

\subsection{Challenge 7: ReRAM Cell Resolution}

The precision of the weights significantly affects the accuracy of the output results. DNN training and inference on conventional GPU platforms are done using 32-bit floating-point precision. In the case of high resolution ReRAM cells, there is a need for fewer crossbar arrays, which benefits in lower latency and better accuracy \citet{kao2022design}. However, ReRAM cells have limited states and suffer from low precision. Prior work demonstrated that a 16-bit-wide fixed-point number representation is adequate for classification problems \citet{gupta2015deep}. Besides, a 4-bit ReRAM cell is more susceptible to temperature variation than a 2-bit ReRAM cell due to its having a larger number of states\citet{joardar2020accured}. Moreover, intermediate states are more vulnerable to heat than the states close to electrodes \citet{shim2021impact}.  As mentioned earlier, an increase in temperature leads to a decrease of the G$_{on}$/G$_{off}$ ratio and lowers the noise margin (NM) \citet{joardar2020accured}.  In particular, utilization of an 8-bit cell instead of a 2-bit cell decreases the number of required resistive crossbar arrays by 75\%, but it also leads to a 64$\times$ NM drop \citet{yang2021multi}. Therefore, numerous ReRAM-based accelerators have adopted a weight-composing scheme \citet{shafiee2016isaac}, \citet{ankit2019puma} with an increased number of arrays and additional power consumption and latency \citet{smagulova2021resistive}. 
\par



\subsection{Challenge 8: Input Distribution}
The amplitude and frequency of input signals contribute to the power density and speed of the hardware, respectively. An increase of input voltages leads to higher power consumption and generates heat \citet{zhang2022thermal}. 
Higher operating frequency decreases execution time, but thermal noise causes limitations in the frequency scaling of ReRAM-based designs \citet{joardar2020accured} and an increase of frequency leads to accuracy decrease  \citet{liu2019hr}, \citet{yang2021multi}. The degradation gets more severe as the depth of neural networks grows. For instance, at 1GHz the accuracy of VGG-19 drops by 50\% compared to 5\% in LeNet-5. Adding residual connections improves accuracy \citet{joardar2020accured}.
At frequency 100MHz and temperatures of 300K and 400K, the accuracy of ResNet20 was equal to around 88.2\% and 88.1\%, respectively. When frequency increased to 1GHz, at the same temperature conditions accuracy decreased to 83.5\% and 80.5\%, respectively \citet{yang2021multi}. 
\par 




\section{Solutions}
\label{sec:solutions}

\begin{figure}[hh]
\centering
\includegraphics[width=0.45\textwidth]{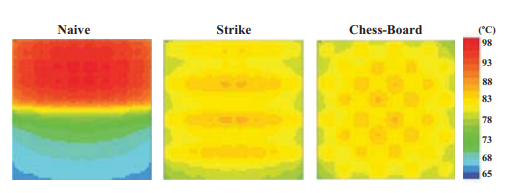}
\caption{ Steady-state temperature distributions of the bottom ReRAM layers: naive, strike and chess-board allocation \citet{zhou2019thermal}. }
\label{conduction}
\end{figure}

In \citet{zhou2019thermal} the authors tested three memory allocation schemes - "strike," "chess-board" and "naive" - in 3D crossbar array for the same model and in all cases the bottom ReRAM layers were the hottest. As can be seen from Figure \ref{conduction}, the peak temperature of the "naive" scheme was as high as 8K, which also demonstrates that the thermal distribution can be controlled to a certain extent by static allocation schemes. Moreover, the temperature distribution in chips might dynamically change with time. Therefore, modern techniques should be able to mitigate the negative impact of high temperatures on ReRAM chips on the fly. 
 
In this section, some of the SoTA solutions involving thermal-aware optimization of ReRAM-based memory and PIM accelerators are presented. 

\subsection{Solution 1: THOR }

The goal of the \textit{thermal-aware optimization for extending ReRAM lifetime (THOR)} is to keep the temperature of the ReRAM banks below a threshold temperature to ensure a lifetime above ESL. THOR consists of THOR - Lazy Access (THOR-LA) and THOR - Smart Access (THOR-SA) schemes, which can work both together and independently from each other. THOR-LA delays requests to hot banks and thus allows their cooling during idle periods. The delays are implemented by extending the memory controller (MC) to four queues: Normal read/write and Lazy read/write. THOR-SA reduces the number of accesses to hot arrays. 
\par

Nevertheless, it allows overall system power reduction by 5.5\%  and ReRAM lifetime enhancement by 2.06$\times$ the baseline design with a normal read queue, a lazy read queue, a normal write queue, and a lazy write queue \citet{beigi2019thermal}.

\subsection{Solution 2: DeepSwapper}

Hybrid DRAM/NVM memory systems benefit from the lower access latency of DRAM and the high capacity of ReRAM. However, data migration between two memory types is costly due to the need for metadata storage.  Existing swapping schemes are based on prediction tables and do not consider the temperature effect. \par 
DeepSwapper is a novel deep learning-based page swap management scheme for hybrid DRAM/ReRAM memory. Instead of lookup tables, it uses a Long Short-Term Memory (LSTM) recurrent neural network (RNN) to predict future memory access patterns. This hardware-managed framework consists of two main components: an LSTM-Based Address Predictor and a Temperature-Aware Swap Management Unit. Evaluation results showed that the ReRAM lifetime was enhanced by 1.87$\times$ that of other schemes \citet{beigi2019thermal}.




\subsection{Solution 3: TADMSIMA
}


\textit{Thermal-Aware Design and Management for Search-based In-Memory Acceleration (TADMSIMA)} is a thermal-aware data allocation scheme that utilizes steady-state and dynamic thermal management (DTM) techniques \citet{zhou2019thermal}.  In the first stage, static program analysis is used to estimate the number of ReRAM banks and their power consumption based on the type of application program, the size of the dataset, the architecture and teh operating frequency. Then, banks are classified as high power-consuming and low power-consuming. For thermal-aware mapping, a two-phase design space exploration method based on a genetic algorithm is applied. 


The proposed system was validated on two search-based applications - hyperdimensional computing and database query processing. The experimental setup included 10 encoding-search ReRAM bank pairs to store and compute data.  According to the results, the steady-state temperature was reduced by at least 15.3K 
and the lifetime of the ReRAM device was  extended by 57.2\% on average. The dynamic temperature management provided 17.6\% performance improvement compared to other SoTA methods.

\begin{figure}[!t]
\centering
     \begin{subfigure}[]{0.3\textwidth}
     \centering
        {\resizebox{0.6\textwidth}{!}{\includegraphics{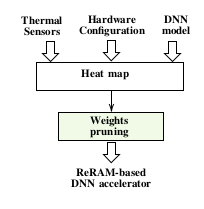}}}
        \vspace{10mm}
        \caption{}   
    \end{subfigure}
    \begin{subfigure}[]{0.3\textwidth}
        {\resizebox{1\textwidth}{!}{\includegraphics{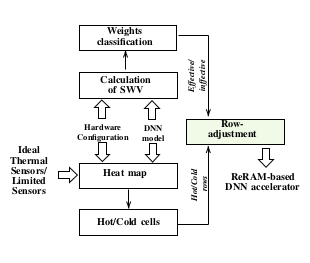}}} 
        \vspace{1mm}
        \caption{}
    \end{subfigure}
    \hspace{3mm}
    \begin{subfigure}[]{0.3\textwidth}
    \centering
       {\resizebox{0.7\textwidth}{!}{\includegraphics{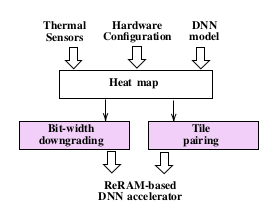}}}
       \caption{}
        \vspace{5mm}
    \end{subfigure}
     \begin{subfigure}[]{0.3\textwidth}
      {\resizebox{0.8\textwidth}{!}{\includegraphics{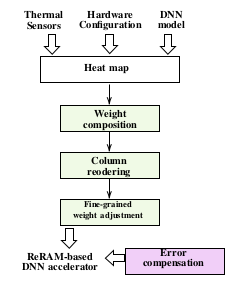}}}
      \vspace{8mm}
      \caption{}
    
    \end{subfigure}
      \begin{subfigure}[]{0.3\textwidth}
        {\resizebox{0.9\textwidth}{!}{\includegraphics{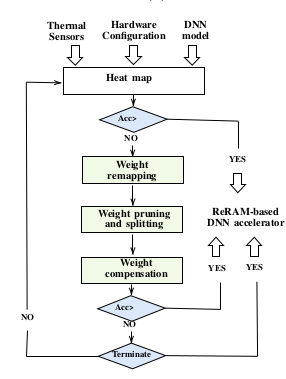}}}
        \caption{}

    \end{subfigure}
    \begin{subfigure}[]{0.3\textwidth}
        {\resizebox{0.9\textwidth}{!}{\includegraphics{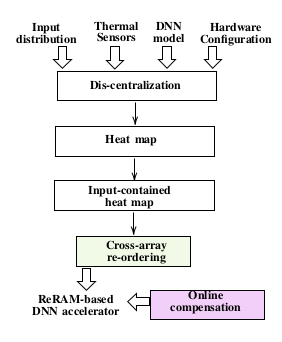}}}
        \vspace{3mm}
        \caption{}
    \end{subfigure}

\caption{Thermal-aware solutions for acceleration of DNN and IMC: a) The Temperature-Aware Weight Adjustment (TAWA) scheme; b) Thermal-aware optimization of ReRAM-based neuromorphic computing systems (TARA); c) A Heat Resilient Design for RRAM-based Neuromorphic Computing (HR$^{3}$AM); d) A Thermal-aware Optimization Framework for Accelerating DNN on ReRAM (TOPAR); e) Weight RemApping and processing in RRAM-based Neural Network Accelerators (WRAP); f) Thermal-aware layout optimization and
mapping methods for resistive neuromorphic engines (TALOMRNE). }
\label{dnn}
\end{figure}

\subsection{Solution 4: TARA}

One of the first works that addressed the impact of temperature on computational accuracy in a ReRAM-based neuromorphic computing system proposed \textit{weight remapping} based on \textit{temperature-aware row adjustment (TARA)} \citet{beigi2018thermal}. Its performance is compared to a baseline architecture with random mapping and an architecture with \textit{temperature-aware weight adjustment (TAWA)}. TAWA is based on weight pruning, as proposed in \citet{han2015learning}, and its implementation scheme is illustrated in Figure \ref{dnn}a. Here, weights mapped to hot cells are pruned and the neural network is retrained again. 
\par 
The schematic of TARA  shown in Figure \ref{dnn}b was designed for Micron’s HMC architecture.  Here, diode thermal sensors were placed at the center and left side of each row of ReRAM array - the hottest spots due to the close location of the analog-to-digital (ADC) converter and memory. At each epoch time, the temperature of the rows was approximated and classified. If the estimated temperature of a row was higher than the threshold temperature equal to 330K,
the ReRAM crossbar row was considered hot; otherwise, cold. In addition, rows of neural networks were classified as \textit{effective} and \textit{ineffective} using a metric called Summed Weight Variations (SWV) and predefined threshold $\beta$:

\begin{equation}
    SWV_{pq} = \sum_{j=0}^{m} |w_{pj}-g_{qj}|
\end{equation}
where $w_{pj}$ is the weight at the location $(p,j)$ and $g_{pj}$ is the corresponding conductance of the ReRAM cell. If $SWV>\beta$, rows are effective; otherwise, ineffective.




\begin{figure}[hh]
\centering
\resizebox{0.4\textwidth}{!}{
    
    \includegraphics{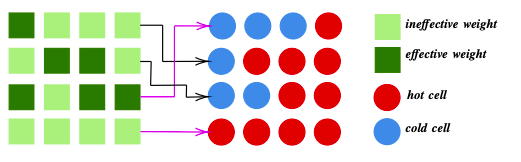}
}

\caption{Weight mapping in TARA.} 
\label{TORNCS1}
\end{figure}
 
Increase of row temperature from 
340K to 360K
leads to an accuracy decrease from 61.7\% to 23.4\%. Therefore, at the final stage of the scheme, effective rows were mapped to ReRAM crossbar array avoiding hot rows, as in Figure \ref{TORNCS1}. Evaluation of the \textit{thermal-aware row adjustment} on a two-layer neural network in NeuroSim demonstrated an increase of the system accuracy by up to 39.2\%.



\subsection{Solution 5: HR$^{3}$AM: A Heat Resilient Design for RRAM-based Neuromorphic Computing}

Conversion of neural network weight $w$ into conductance state of ReRAM cell $G$ can be done based on the equation below:
\begin{equation}
   G = \alpha \times w + \beta
\end{equation}
where parameter $\alpha = \frac{G_{max}-G_{min}}{w_{max}-w_{min}}$ is used to scale a weight $w$ within a range of [$G_{min},G_{max}$] and parameter $\beta = G_{max}- \alpha \times w_{max}$ is used to remove negative weights. 
\par 
It was observed that a 1$^{\circ}$ increase of temperature in ReRAM-based architecture leads to an overall performance decrease of 0.9\%. In order to decrease the negative impact of heat on ReRAM-based CNN accelerators, the HR$^{3}$AM design (Figure \ref{dnn}c) utilizes a  \textit{bitwidth downgrading} technique (HR$^3$AM-BD) and \textit{tile pairing} (HR$^3$AM-TP) \citet{liu2019hr}. 
To do this, the HR$^3$AM system monitors temperature distribution in the ReRAM chip dynamically using temperature sensors. If the temperature is above the threshold (330K), a heat-resilient weight adjustment is applied:
\begin{equation}
    G_{new} =\frac{1}{2^{N}}\times (\alpha
\times w + \beta)\end{equation}
where $G_{new}$ is the new conductance state and $N$ is the number of shifted bits so that:
\begin{equation}
V_{o} = V^{T}_{i}\times G_{new}\times R_{S}\times 2^{N} = \\ (V^{T}_{i}\times G_{old}\times R_{S}/2^{N})\times 2^{N}.    
\end{equation}
where $V_{o}$ is the output voltage; $V_{i}$ is the input voltage; and $G_{old}$ is the old weight. \par
In addition to this HR$^3$AM-BD, thermal distribution can be reduced by the introduction of \textit{master} and \textit{slave} tiles. The overheated (master) tile  is paired with a cooled-down idle (slave) tile in such a way that the output of the master tile is read from even-index columns $V_{out}^{m}$ = \{$v_{0}, v_{2}...v_{2N}$\} and the output of the slave tile is read from odd-index columns $V_{out}^{s}$ = \{$v_{1}, v_{3}...v_{2N+1}$\}. This decreases the number of functioning cells in a crossbar array and thus reduces power consumption. The pairing mode is represented by \textit{a pairing bit} and \textit{a master/slave bit} in crossbar arrays.  

The design was tested on a small two-layer network for MNIST classification and larger networks such as VGG16, ResNet50 and InceptionV3 for ImageNet classification. The obtained results showed 4.8\%-58\% improvement compared to the baseline model, which has no thermal optimization. In addition, HR$^{3}$AM showed better accuracy by 4.3\%–41.8\% over TARA \citet{beigi2018thermal}. 

\subsection{Solution 6: TOPAR}

To reduce average temperature and temperature variance between ReRAM arrays in DNN accelerators, a \textit{thermal-aware optimization framework for accelerating DNN on ReRAM (TOPAR)} has been proposed \citet{shin2020thermal}. It consists of three-stage offline thermal optimization and online thermal-aware error compensation, as shown in Figure \ref{dnn}d. 
\par 
There are $2^N-V$ ways to decompose an N-bit  weight value $V$ to positive and negative arrays. To reduce the temperature in the ReRAM chip, the first step of the offline stage performs a thermal-aware weight decomposition (TOPAR-I).  In other words, TOPAR-I aims to identify a decomposition case with the smallest sum of partial weights. The next step,  a thermal-aware column reordering (TOPAR-II), shuffles the order of the column pairs in positive and negative ReRAM arrays. This changes to weight and temperature distribution in ReRAM arrays and does not affect the computational output.   The final step in offline optimization (TOPAR-III) is a fine-grained weight adjustment if there are more than two decomposition cases in TOPAR-I. It is performed sequentially starting from the top-left position of the crossbar array. TOPAR-III aims to reduce the cost difference between positive and negative arrays. At online stage,
TOPAR improves ReRAM endurance by up to 2.39$\times$ and preserves inference accuracy.


\begin{figure*}[!t]
\centering
\begin{minipage}[t]{.2\textwidth}
        \centering
        \resizebox{1\textwidth}{!}{\includegraphics{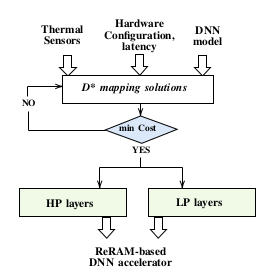}}
        \vspace{1mm}
        \subcaption{}\label{fig:11}
    \end{minipage}
    \begin{minipage}[t]{.2\textwidth}
        \centering
        \resizebox{0.55\textwidth}{!}{\includegraphics{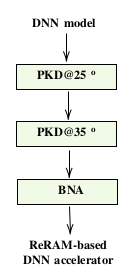}}
        \subcaption{}\label{fig:12}
    \end{minipage}
    \begin{minipage}[t]{.55\textwidth}
        \centering
        \resizebox{0.9\textwidth}{!}{\includegraphics{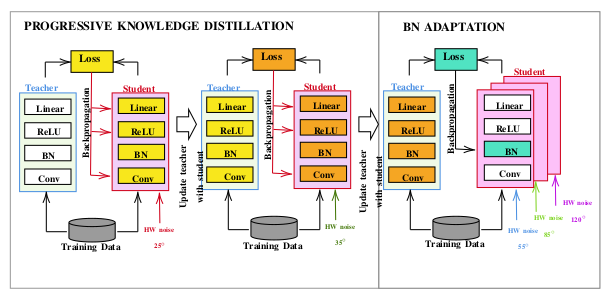}}
        \subcaption{}\label{fig:2}
    \end{minipage}

\caption{ a) Placement strategy in AccuReD; b)  Temperature-Resilient ReRAM-based In-Memory Computing for DNN Inference (TRRIMC); c) Implementation of PKD and BNA in TRRIMC. } 
\label{fig:M1}
\end{figure*}

\subsection{Solution 7: Weight RemApping and Processing in RRAM-based Neural Network Accelerators (WRAP)}

A weight remapping and processing (WRAP) framework adopted the \textit{subarray-based} approach rather than dealing with each weight individually \citet{chen2022wrap}. This helped to reduce computational complexity accuracy while mitigating thermal issues to maintain the system. Figure \ref{dnn}e shows the flow of WRAP which is based on three algorithms:  weight remapping (WR); weight pruning and splitting (WPS); and weight compensation (WC).

At the initial stage, the framework receives parameters of DNN model hardware such as ReRAM cell resolution and ReRAM array size and maps weights to the accelerator. Afterward, it retrieves a heatmap of layers and estimates the accuracy of the system \textit{Acc}. If the latter is below a predefined threshold level $\theta$, three subarray-based algorithms are applied to remap the weights until the estimated accuracy is above the threshold accuracy. In addition, it was observed that deep layers are less sensitive to pruning than shallow layers. The process is terminated when the "prune ratio" in WPS is zero.

The framework was evaluated on VGG8, VGG11, ResNet34 and AlexNet for CIFAR-10 classification with less than 2\% inference accuracy loss  \citet{chen2022wrap}.

\subsection{Solution 8: TALOMRNE} 
\label{subsec:8}

\textit{Thermal-aware layout optimization and mapping methods for resistive neuromorphic engines (TALOMRNE)} 
introduced a new layout that implies decreased temperature distribution by dis-centralizing components of the accelerator. In addition, TALOMRNE (Figure \ref{dnn}f) also noted that previous works emphasize only the weight itself and do not consider the input distribution that contributes to the final power consumption of the system. In addition, it adopted the cross-array swapping method of input-contained weights $W_{bias}$. 



The method validation was done for two SNN models transformed from VGG9 and VGG11 on CIFAR-10 and CIFAR-100 datasets, respectively. The method allowed reducing the peak temperature up to 10.4$^{\circ}$ and improved the endurance
by up to 1.72$\times$ \citet{zhang2022thermal}.

\subsection{ Solution 9:  AccuReD } 
\textit{AccuRed} is a heterogeneous ReRAM-GPU-based architecture for CNN training and inference.  Its compute-intensive layers are mapped to ReRAM arrays, whereas precision-critical layers are mapped on GPUs. In addition, AccuRed performs a thermal-aware placement strategy (Figure \ref{fig:M1}a) based on a joint performance–thermal-aware mapping and a thermal reference cell (TRC) to reduce temperature impact.
For effective mapping and high accuracy, AccuReD applies the Multiobjective Optimization (MOO) technique. The combined objective can be expressed as follows:
\begin{equation}
    D^\ast = MOO(D, OBJ = {U(d), T(d)}) 
\end{equation}
where $D^\ast$ is the set of Pareto optimal mapping solutions among CNN layer mapping $D$; $d$ is candidate mapping; $T$ is the temperature objective combining both horizontal and vertical heat flow models; $MOO$ is a multiobjective optimization solver; and $OBJ$ is the set of all objectives (latency and temperature). AMOSA \citet{bandyopadhyay2008simulated} was used as the MOO solver to minimize \textit{cost}, which is a function of latency and temperature. Performance- and thermal-aware mapping of $L$ layers of CNN to $N$ processing elements (GPUs and ReRAMs) ensures low  temperature and high performance. The CNN layers are classified as high-power (HP) and low-power (LP) layers. During one pipeline stage, the HP layer requires the ReRAM to be active for more than 50\% and placed near the sink. LP layers can be placed farther from the sink. Comparison of TSV-based and M3D designs showed that M3D integration of AccuRed has superior thermal characteristics and allows more CNN layers without sacrificing accuracy. AccuRed outperformed conventional GPUs by 12$\times$ on average and can be further scaled up \citet{joardar2020accured}.

\subsection{Solution 10: TRRIMC }

Noise injection during DNN model training is one of the ways to increase the robustness of ReRAM-based accelerators when it comes to temperature variations. In addition, knowledge distillation \citet{hinton2015distilling} can improve the performance of the model accelerated on the hardware. However, such a model recovers only in certain conditions and fails in other scenarios. To improve the generality of the model, authors  \citet{meng2021temperature} proposed a novel training algorithm 
- progressive knowledge distillation (PKD) - and thermal-aware batch normalization adaptation (BNA). The schematic of the \textit{Temperature-Resilient RRAM-based In-Memory Computing for DNN Inference (TRRIMC)} is shown in Figures \ref{fig:M1}c and \ref{fig:M1}d.

\par 
In PKD, a clean low-precision model is used as a teacher model. In the initial phase, a student model, duplicated from a teacher model, is trained with low-temperature noises. Then, the trained model acts as a new teacher model and a new duplicated student model is trained with higher temperature noises and so on. During BNA, weights and learnable parameters of DNN are frozen and further training of 16-bit fixed-point batch normalization (BN) parameters $Y_{BNA}$ with noise injection at different temperature $T$ scenarios is performed:  
\begin{equation}
Y_{BNA} = w_{T} \times \frac{Y-\mu_{T}}{\sigma_{T}} + b_{T}
\end{equation}
where $Y$ is the output preactivation; $w$ is the weight; $\mu$ is the mean within the batch and $\sigma$ is its standard deviation. 
\par
The proposed PKD+BNA method allows recovering the accuracy of the 2-bit ResNet on the CIFAR-10 for more than 30\% and of the 4-bit ResNet-18 on TinyImageNet for more than 60\%. The primary advantage of the scheme is the absence of the need to reprogram the initial ReRAM weights.

\section{Discussion and Takeaways}
\label{sec:discussion}

The nanoscale size and non-volatile nature of RSMs  allow implementation of small-size and energy-efficient computational hardware components based on ReRAM crossbar arrays. On the other hand, dense architecture design increases temperature susceptibility and negatively affects reliability since an active ReRAM cell in a crossbar array causes thermal disturbance (TBD) in neighboring victim cells \citet{sun2015thermal}. Therefore, the accuracy of a ReRAM simulation model plays a vital role in the validation of the ReRAM-based hardware design before its fabrication. One of the ways to overcome the conductance drift problem in RCAs is to frequently refresh the cells' states but it requires additional power consumption \citet{meng2021temperature}. Other solutions were designed to reduce the overall thermal density of the ReRAM-based architecture and were presented in Section \ref{sec:solutions}.

Each of them aimed to overcome certain challenges discussed in Section \ref{sec:challenges} such as the recovery of computation accuracy, extending the lifetime of ReRAM cells, and reduction of power consumption. Nevertheless, solving one of these challenges also leads to improvement in other challenges. Table \ref{solutions} lists the proposed solutions for ReRAM-based memory devices and accelerators.  In particular, \textit{Solutions 1-3} in Section \ref{sec:solutions} are designed for NVM memory architectures. They aim to achieve a thermal-optimal data allocation by implementing  a memory access control in THOR, and utilization of neural network and genetic algorithm in DeepSwapper and TADMSIMA, respectively. In addition, access control can be done. This includes introduction of idle periods allowing ReRAM cells to cool down. The main aim of \textit{Solutions 4-8}  in Section \ref{sec:solutions} is recovery of the accuracy of the system during the acceleration of DNN and CIM tasks. 

Early works considered mainly steady-state temperature distribution cases, whereas recent works propose methods to control runtime temperature variations too.  
The proposed temperature-adjustment schemes for RCAs can be divided into two categories:
\begin{itemize}
    \item temperature-aware optimization and remapping;
    \item temperature-resilient training of the DNN model.
\end{itemize} 

The goal of thermal-aware optimization and remapping is to mitigate the impact of high temperature and to create a uniform temperature distribution in a ReRAM crossbar array. The initial stage of these solutions requires the creation of a thermal profile, typically obtained from the limited number of temperature sensors located around ReRAM arrays. In HMC it is the center and left side of the rows\citet{beigi2018thermal}. Afterward,  various weight mapping optimization techniques are applied. These methods are provided in Table \ref{methods}. The offline stage involves temperature-aware training and/or optimization  steps prior to deployment on the ReRAM-based hardware. The online stage includes measures designed to react to dynamic changes of temperature. The optimization and remapping take place at different levels of granularity:  weight level, group of weights (row-/column-wise) level, subarray level, array level and tile level.

\begin{landscape}
\begin{table}[]

\caption{ Thermal-aware ReRAM layout optimization solutions and the challenges they are designed for. }
\resizebox{1\linewidth}{!}
{
\begin{tabular}{|c|c|c|c|c|c|c|c|c|c|c|l|l|}
\hline
\textbf{}   \textbf{Solution (Year)}               & \textbf{Application}                                                 & \textbf{Architecture}                                        & \textbf{Ch1} & \textbf{Ch2} & \textbf{Ch3} & \textbf{Ch4} & \textbf{Ch5} & \textbf{Ch6} & \textbf{Ch7}  & \textbf{Ch8}   &  \textbf{Description}& \textbf{Setup and Tools} \\ \hline

\textbf{\begin{tabular}[c]{@{}c@{}} THOR \\ (2018)\end{tabular}}        &     NVM memory   & \begin{tabular}[c]{@{}l@{}}2.5D/3D\\ interposer\end{tabular} & &              &     \cmark          &        &              &              &              &          & \begin{tabular}[c]{@{}l@{}}THOR-LA has four queues: Normal read, Lazy read, Normal write, and Lazy write. \\ 1. ReRAM banks are classified as hot and cold banks based on the sensed temperature. \\ 2. Read and write requests to hot banks are delayed, allowing to cool them.\\ Lifetime enhancement by 1.70$times$. Power reduction by 6.7\%. \\
THOR-SA has two queues: Normal read, Normal write.\\ 1. Six bits are added to cache tags (two bits to show rank ID and four bits to identify bank number).\\ 2. LLC are sampled to sets with hot and cold banks and maintained in the least recently used (LRU) order.\\ 3. Two hit counters: hot hit counter and cold hit counter. \\ 4. Maintains a temperature-aware policy to keep cache lines from hot banks longer in the LLC \\ and reduce the number of future accesses to hot banks.\\ Lifetime enhancement by 1.36$times$. Power reduction by 4.6\%.\end{tabular} & 
\begin{tabular}[l]{@{}l@{}}1. gem5 simulator integrated \\ with NVMAIN + \\ CACTI +
DESTINY \\ 2. Requires additional hardware\end{tabular} \\ \hline

\textbf{\begin{tabular}[c]{@{}c@{}}DeepSwapper\\ (2019)\end{tabular}} & \begin{tabular}[c]{@{}c@{}}Hybrid \\ NVM/DRAM \\ memory\end{tabular} & N/A                                                             &              &             &   \cmark       &              &              &              &         &  & \begin{tabular}[c]{@{}l@{}}1. Seq2seq LSTM: A sequence of past LLC miss addresses is used to predict a sequence of future LLC\\  miss addresses. \\ 2. A beam-search decoder is used to improve LSTM. \\ Endurance improvement by 1.87$\times$. \end{tabular} & \begin{tabular}[l]{@{}l@{}}1.gem5 with Ramulator \\ 2. A two-layer depth LSTM model \\ with 128 and 64 hidden units \\ 3. P100 NVIDIA GPU \end{tabular}  \\ \hline
\textbf{\begin{tabular}[c]{@{}c@{}}TADMSIMA \\(2019) \end{tabular}}    &   \begin{tabular}[c]{@{}c@{}}    Search-based \\ hyperdimensional\\ computing \\ and database query \\ processing  \end{tabular}                                                              &     HMC-like                                                         &  \cmark            &              &    \cmark          &              &              &              &         &   & \begin{tabular}[l]{@{}l@{}} 1. Static program analysis is used to estimate the number of ReRAM banks  and their power \\consumption based on the type of application program, the size of the dataset, the architecture \\ and the operating frequency. \\2. Banks are classified as high power-consuming and low power-consuming. \\
3. For thermal-aware mapping, a two-phase design space exploration method based on genetic \\ algorithm is applied. \\
Steady-state temperature reduction by at least {\color{red}15.3$^{\circ}$} and ReRAM lifetime enhancement by 57.2\% on average.  \end{tabular} & \begin{tabular}[l]{@{}l@{}} 1. McPAT \\2. CACTI \\  3. HSPICE\\4. HotSpot \\ 5. Ni/HfO$_{2}$/PT and Ti/TiO$_{2}$/Pt\end{tabular}  \\ \hline

\textbf{\begin{tabular}[c]{@{}c@{}}TARA \\(2018)\end{tabular}}      & \begin{tabular}[c]{@{}c@{}}CIM/DNN \\ inference \\(a 2-layer NN on\\ MNIST ) \end{tabular}                                                                  & HMC                                                          &   \cmark     &              &              &                &              &              &         &  & \begin{tabular}[l]{@{}l@{}}Baselines: neuromorphic hardware with random mapping scheme (accuracy 54.1\%); TAWA (accuracy 62\%). \\ 1. Temperature collection and estimation. \\ 2. Classification NN weight rows to effective and ineffective based on SWV.\\3. Temperature-aware row adjustment, e.g. avoiding mapping effective rows to hot ReRAM cells. \\
Accuracy improvement by 23.8\% compared to the baseline;  by 14.3\% more than the TAWA scheme.\end{tabular} & \begin{tabular}[l]{@{}l@{}}1. 1-bit per ReRAM \\2. [$G_{off};G_{on}$]=[3.07nS, 38.4nS] 
\\3. NeuroSim \end{tabular}     \\ \hline
\textbf{\begin{tabular}[c]{@{}c@{}}HR$^{3}$AM \\ (2019)\end{tabular}}       &  \begin{tabular}[c]{@{}c@{}}CNN inference \\(1. a small two-layer\\ NN on MNIST \\ 2. VGG16, ResNet50 \\ and InceptionV3  \\on ImageNet) \end{tabular}                                                                     & HMC                                                              &   \cmark           &              &              &              &              &              &         &    &  \begin{tabular}[l]{@{}l@{}}Baseline: neuromorphic hardware with random mapping scheme, TARA. \\ 1. HR$^{3}$AM-BD: temperature-aware row adjustment using bit-width downgrading technique.\\ HR$^{3}$AM-BD  aims to improve accuracy.  (Suitable for DNN with large cell resolution.) \\2. HR$^{3}$AM-TP: the tail pairing method is when some operations from hot tiles are performed on the idle tile. \\HR$^{3}$AM-TP aims to reduce the temperature of the chip.  (But extra tiles lead to a loss in parallelism.) \\  Decrease in maximum temperature by 6.2$^{\circ}$; decrease in average temperature for the entire chip by 6$^{\circ}$; \\ Accuracy improvement by 4.8\%–58\% over the baseline and
4.3\%–41.8\% over TARA.

\end{tabular}
&  \begin{tabular}[l]{@{}l@{}}1. Based on ISAAC \\2. Requires temperature sensors,\\ registers and comparators \\3. A downgrade bit and control logic \\4. Adjustment of shift-and-add \\ and encoding circuits \\5.  Reserved idle tiles (10\% of all tiles) \\ 6. Tensorflow \end{tabular}\\ \hline
\textbf{\begin{tabular}[c]{@{}c@{}}TOPAR \\ (2020)\end{tabular}}       &  \begin{tabular}[c]{@{}c@{}}DNN inference \\ (ResNet18, ResNet50, \\ VGG-16, neural \\ collaborative \\filtering (NCF), \\ a 2-layer stacked\\ LSTM) \end{tabular}                                                                      &     N/A                                                         &              &              &      \cmark        &              &              &              &      &     & \begin{tabular}[l]{@{}l@{}}Baseline: HR$^{3}$AM \\ Offline optimization:\\ 1. TOPAR-I: a thermal-aware weight decomposition. \\ 2. TOPAR-II: a thermal-aware column reordering. \\ 3. TOPAR-III: a fine-grained weight adjustment.  \\ Online optimization:\\ 1. Restoring distorted current-sum results with the thermal-aware error compensation.\\ Improved endurance up to 2.39$times$.\end{tabular}  & \begin{tabular}[l]{@{}l@{}}1. Based on ISAAC \\ 2. 2-bit ReRAM cell \\ 3. 64$\times$64 array \\4.  HotSpot thermal simulator \\ 5. Pytorch \\ 6. 8-bit weights \\ 7. Endurance 4.14$\times$10$^8$ \\ 8. Synopsys Design Compiler \end{tabular}   \\ \hline

\textbf{\begin{tabular}[c]{@{}c@{}}WRAP\\(2022)\end{tabular}}        &  \begin{tabular}[c]{@{}c@{}}CIM/DNN\\  inference \\ (VGG 8, VGG11, \\ Alexnet and\\ ResNet34)\end{tabular}    &   \begin{tabular}[c]{@{}c@{}} 3D \\ (4-layer \\ HMC-like) \end{tabular}                                                       &  \cmark              &              &              &              &              &              &        &   & \begin{tabular}[l]{@{}l@{}} Baseline: HR$^{3}$AM \\ Subarray-based approach saves computational resources. \\ 1. WR: avoiding mapping of important weights to higher temperature subarrays.
\\2. WPS: less critical weights are pruned, which frees some subarrays.
\\3. WC:  bitwidth downgrading on subarrays.\\
Accuracy loss is less than 2\%;  and less than 1\% loss with compensation. 
\end{tabular}  & \begin{tabular}[c]{@{}l@{}} 1. Based on ISAAC \\ 2. [$G_{off};G_{on}$]=[3.07nS, 38.4nS] \\ 3. Pytorch \\4. 4-bit, 6-bit, 8-bit weights \\5. Pruning ratio 40–50\% (best results) \end{tabular}  \\ \hline

\textbf{\begin{tabular}[c]{@{}c@{}}TALOMRNE \\(2022)\end{tabular}}    &    SNN                                                                 &  N/A                                                            &              &              &    \cmark          &              &              &              &        &    \cmark  & \begin{tabular}[l]{@{}l@{}} Takes into consideration input distribution and utilizes a cross-array mapping method.\\
1. Layout optimization by dis-centralizing high-density components. \\
2. Thermal-aware weight reordering considering input distribution and weights value.  \\
Average power range decreases by 20\% in \textit{Conv} and by 15\% in \textit{FC} layers.  \\
Endurance improvement by 1.30$\times$ and 1.72$\times$ in VGG-11 and VGG-9, respectively.
\end{tabular} & \begin{tabular}[l]{@{}l@{}}1.Based on ISAAC\\
2. Block size 128$\times$128 \\ 3. Input voltage [0;0.9]V \\ 4. 
 [R$_{on}$;R$_{off}$]=[5k$\Omega$; 500k$\Omega$] \\ 5. Hotspot \\  6. Endurance 4.14$\times$10$^8$ \end{tabular}\\ \hline
\textbf{\begin{tabular}[c]{@{}c@{}}AccuReD \\ (2020)\end{tabular}}     & \begin{tabular}[c]{@{}c@{}} Hybrid NVM/GPU; \\ CNN training/\\ inference \\ (VGG-19)\end{tabular}    & M3D                                                              &              &              &              &              &              &              &          \cmark &     & \begin{tabular}[l]{@{}l@{}}  
Aims to reduce \textit{cost}, which is a function of temperature and CNN pipeline latency. \\
1. Used a thermal reference cell (TRC) and multicell reference array. \\ 2. MOO and thermal-aware mapping. AMOSA  used as the MOO solver.  \end{tabular} & 
\begin{tabular}[l]{@{}l@{}} 1. Based on AccuReD \\ 
2. GPGPU-Sim \\3. PytorX \end{tabular} \\ \hline


\textbf{\begin{tabular}[c]{@{}c@{}}TRRIMC \\ (2021)\end{tabular}}      & \begin{tabular}[c]{@{}c@{}}CIM/DNN\\ training \\ (a 2-bit ResNet-18 on \\CIFAR-10 and \\ TinyImageNet)  \end{tabular}        &      N/A                                                        &              &              &              &              &              &              &         &  &  \begin{tabular}[l]{@{}l@{}} Basline: 2-bit ResNet-18 on the CIFAR-10 with perodic refreshing after 30 $s$ of operation.\\
1. PKD training. \\ 2. Thermal-aware BNA. \\ Accuracy is > 90\%  until 10$^{4}$ $s$ with reduced refreshing frequency by 250$\times$.\end{tabular} & \begin{tabular}[l]{@{}l@{}} 1. 90nm prototype chip \\ 2. 2-bit $HfO_{2}$ ReRAM cell \\ 3. 256$\times$256 ReRAM array \\ 4. NeuroSim \\ 5. For BNA: temperature sensors \\and BN multiplexer   \end{tabular}     \\ \hline

\multicolumn{1}{l}{Ch: Challenge; N/A: not reported} 
\end{tabular}
}
\label{solutions}
\end{table}
\end{landscape}

\begin{landscape}
\begin{table}[]
\centering
\caption{Methods applied to decrease temperature effect in ReRAM-based CIM accelerators }
\resizebox{1\linewidth}{!}
{
\begin{tabular}{|c|l|l|c|c|c|c|l|l|}
\hline

\textbf{Approach}                                                                                                    & \multicolumn{1}{c|}{\textbf{Method}}                                                                   & \multicolumn{1}{c|}{\textbf{Implementation}} &
\multicolumn{1}{c|}{\textbf{Level}} &
\multicolumn{1}{c|}{\textbf{Solution}} &
\multicolumn{1}{c|}{\textbf{Offline}}
&
\multicolumn{1}{c|}{\textbf{Online}}
& \multicolumn{1}{c|}{\textbf{Advantages}} & \multicolumn{1}{c|}{\textbf{Disadvantages}} \\ \hline

                          \multirow{9}{*}{\textbf{\begin{tabular}[c]{@{}c@{}}Temperature-\\ aware \\optimization and  \\weight remapping
\end{tabular}
}
} 

&
Weight pruning (WP)                                                                       &   \begin{tabular}[c]{@{}l@{}}pruning weights of hot ReRAM \\ cells and retraining of DNN \end{tabular}            &    \begin{tabular}[c]{@{}l@{}}array\end{tabular}            &    \begin{tabular}[c]{@{}l@{}}TAWA\end{tabular}                        & \cmark &\xmark                &  \begin{tabular}[c]{@{}l@{}}hot ReRAM cells are \\ excluded from utilization  \end{tabular}                                       &   \begin{tabular}[c]{@{}l@{}} requires extra training \\  \end{tabular}                                          \\ \cline{2-9} 
& \begin{tabular}[c]{@{}l@{}}Row adjustment  (using the \\ $SWV$ metric)\end{tabular} & \begin{tabular}[c]{@{}l@{}} effective weights are mapped \\ to cold ReRAM array rows \end{tabular}  & array   & TARA  & \cmark &\xmark                               & \begin{tabular}[c]{@{}l@{}} row level; \\ able to recover accuracy of \\ DNN model \end{tabular}                                        &  \begin{tabular}[c]{@{}l@{}} the ambient temperature might increase \\ and degrade performance \end{tabular}                                        \\ \cline{2-9} 
 & Bit-width downgrading                                                                                 &  \begin{tabular}[c]{@{}l@{}}  weight is shifted from a \\ temperature sensitive conductance \\ state  to a conductance state with \\less sensitivity and the obtained \\ multiplication result is shifted \\ back
 \end{tabular} &   weight &   HR$^3$AM   & \cmark &\cmark                                  &      \begin{tabular}[c]{@{}l@{}} reduces thermal effect on \\ weights with high conductance  \end{tabular}                                    &  \begin{tabular}[c]{@{}l@{}}cannot be used for weights with low \\ resolution \end{tabular}                                           \\ \cline{2-9} 

& \begin{tabular}[c]{@{}l@{}}Tile pairing\end{tabular} &   \begin{tabular}[c]{@{}l@{}} pairing of overheated and \\ cooled-down idle tiles   \end{tabular} & array
& HR$^3$AM   & \cmark &\cmark      &  \begin{tabular}[c]{@{}l@{}} reduces the average \\ temperature  \end{tabular}                                          &      requires extra tiles/crossbar arrays                                                                                     \\ \cline{2-9}

 & \begin{tabular}[c]{@{}l@{}} Weight decomposition \\ (WD) \end{tabular}                                                                                   & \begin{tabular}[c]{@{}l@{}} searches for the smallest sum of \\ partial weights among (2$^N$ - V) \\ cases for $N$ bit value V \end{tabular} & weight & TOPAR   & \cmark &\xmark                                    &  \begin{tabular}[c]{@{}l@{}} no additional training required; \\  finding an efficient way of weight \\ decomposition and mapping \\ into positive and negative arrays       \end{tabular}                                 &   \begin{tabular}[c]{@{}l@{}} possible temperature variance between \\ negative and positive arrays \end{tabular}                                          \\ \cline{2-9} 
 
 & Column reordering                                                                                    & \begin{tabular}[c]{@{}l@{}} shuffling order of columns without \\ affecting the computational output \end{tabular} & array & TOPAR                       & \cmark &\xmark                & \begin{tabular}[c]{@{}l@{}} changes distribution of weights \\in an array  and the temperature \\ variance between them \end{tabular}                                         &   \begin{tabular}[c]{@{}l@{}} the process is complicated due to\\ a group of positive and negative \\ arrays \end{tabular}                                          \\ \cline{2-9} 
  &\begin{tabular}[c]{@{}l@{}}  Fine-grained weight \\ adjustment  \end{tabular}                                                                                  &\begin{tabular}[c]{@{}l@{}} WD with  minimum cost upon \\ the result of column reordering; \\performed sequentially \\ \end{tabular} & array & TOPAR                  & \cmark &\xmark                     & \begin{tabular}[c]{@{}l@{}} reduces temperature variation  \\ between positive and negative \\ arrays  \end{tabular}                                         &   \begin{tabular}[c]{@{}l@{}}  limited to weights that have more than \\ two thermal-optimized decomposition \\cases\end{tabular}                                          \\ \cline{2-9} 
& Error compensation & \begin{tabular}[c]{@{}l@{}} current mirror circuits are used \\ for compensation of the current- \\ sum results  \end{tabular} & column &   TOPAR       & \xmark &\cmark                             &                                       \begin{tabular}[c]{@{}l@{}} helps to restore the accuracy  \\ dropped due to the rise of \\ ambient temperature \end{tabular}  &  \begin{tabular}[c]{@{}l@{}} additional area for transistors \end{tabular}                                           \\ \cline{2-9}  
& \begin{tabular}[c]{@{}l@{}}Weight remapping (WR) \end{tabular} & \begin{tabular}[c]{@{}l@{}} avoiding mapping important \\weights of shallow layers \\ to hot subarrays \end{tabular} & subarray & WRAP & \cmark &\xmark                                & \begin{tabular}[c]{@{}l@{}} subarray level; \\
able to recover accuracy of \\ DNN model  \end{tabular}                                        &  \begin{tabular}[c]{@{}l@{}} the ambient temperature might increase \\ and degrade performance \end{tabular}                                        \\ \cline{2-9} 
&
\begin{tabular}[c]{@{}l@{}}Weight pruning  and \\ splitting (WPS)  \end{tabular}                                                                     & \begin{tabular}[c]{@{}l@{}} less-critical (ineffective) weights \\ are pruned and critical (effective)\\ weights are mapped to unused \\ subarrays \end{tabular}                 & subarray      &    \begin{tabular}[c]{@{}l@{}} WRAP \end{tabular}                                      & \cmark &\xmark  & \begin{tabular}[c]{@{}l@{}}  pruning frees several \\ subarrays for mapping \\ critical weights \end{tabular}                                      &  \begin{tabular}[c]{@{}l@{}}"prune ratio" could become 0 before  \\ accuracy reaches above the threshold \end{tabular}                                          \\ \cline{2-9}

& Weigh compensation (WC)                                                                                 & \begin{tabular}[c]{@{}l@{}} weight is shifted from a \\ temperature sensitive conductance \\ state  to a conductance state with \\less sensitivity and the obtained \\ multiplication result is shifted \\ back \end{tabular} & weight &   WRAP   & \cmark &\xmark                                 & \begin{tabular}[c]{@{}l@{}} reduces thermal effect on \\ weights with high conductance  \end{tabular}                                         & \begin{tabular}[c]{@{}l@{}} the compensated result is close to the \\ original \end{tabular} \begin{tabular}[c]{@{}l@{}}  \end{tabular}                                           \\ \cline{2-9}

 & Cross-array reordering                                                                                 & \begin{tabular}[c]{@{}l@{}} weight reordering with \\
 consideration of input signal  \\and weight value \end{tabular} &  cross-array   &   TALOMRNE         & \cmark &\xmark                           &  \begin{tabular}[c]{@{}l@{}} considers input distribution; has \\ larger solutions space  since re- \\ ordering is performed on rows\\ or columns between arrays \end{tabular}                                         &  \begin{tabular}[c]{@{}l@{}} typically requires many iterations \end{tabular}                                           \\ \cline{2-9}

&\begin{tabular}[c]{@{}l@{}} Performance- and thermal-\\aware mapping   \end{tabular}                                                                            & \begin{tabular}[c]{@{}l@{}} CNN layers are classified as high \\power (HP) and low power (LP) \\ HP are mapped near the sink; \\ LP are mapped farther from the sink  \end{tabular}  &  array   &  AccuReD        &  \cmark &  \xmark                         &  \begin{tabular}[c]{@{}l@{}}  implementing multiply-and-\\ accumulate MAC operations \\on ReRAM rather than GPU is more \\ energy-efficient \end{tabular}                                         &  \begin{tabular}[c]{@{}l@{}} exploration of the best mapping in \\ the mapping space can be time- \\consuming \end{tabular}                                           \\ \hline

\multirow{2}{*}{\textbf{\begin{tabular}[c]{@{}c@{}}Temperature-\\ resilient   training\end{tabular}}}               


& \begin{tabular}[c]{@{}l@{}}Progressive knowledge \\ distillation  (PKD) \end{tabular}                                                                                                 & \begin{tabular}[c]{@{}l@{}} aims to minimize discrepancy \\ between teacher and subsequent \\student models trained at \\ different levels of temperature\\ and noise \end{tabular}  &   array    &    TRRIMC                                  & \cmark &\xmark  & \begin{tabular}[c]{@{}l@{}} the model is trained to be \\ robust when it comes to static and dynamic\\ temperature fluctuations \end{tabular}                                        &   \begin{tabular}[c]{@{}l@{}} typically for a short operating time; \\ can lead to overall accuracy degradation  \end{tabular}                                          \\ \cline{2-9} 
       
& \begin{tabular}[c]{@{}l@{}}Batch normalization \\ adaptation (BNA)   \end{tabular}                                                                                            & \begin{tabular}[c]{@{}l@{}} training of batch normalization \\ (BN) parameters with noise-\\ injection  while the weight and \\ learnable parameters  of the NN \\  remain the same  \end{tabular}   &   array    &   TRRIMC                  & \cmark &\xmark                       &  \begin{tabular}[c]{@{}l@{}} improves robustness when it comes to  \\ temperature and hardware \\ compatibility \end{tabular}                                        &    \begin{tabular}[c]{@{}l@{}} adds extra BN parameters \end{tabular}                                         \\   \hline
\end{tabular}
}
\label{methods}
\end{table}
\end{landscape}

One of the basic methods of temperature-aware optimization and remapping  is \textit{weight pruning (WP)}. WP can be applied on either effective or ineffective  weights. In \citet{beigi2018thermal} effective weights that were mapped to hot cells are pruned and NN retrained again. This avoids critical weights being mapped to hot ReRAM cells and maintains accuracy. In \citet{chen2022wrap} ineffective weights are pruned to free space in the arrays. Then, critical weights are remapped. Similarly  \citet{liu2019hr} utilizes the \textit{tile pairing} method to split weights into two tiles. Pairing hot tiles with idle tiles decreases the average temperature since both of them work in low-power mode. 
The next way to change temperature distribution in ReRAM crossbar arrays is to \textit{swap rows or columns} within the same arrays ("in-array") or between arrays ("cross-array") in order to decrease temperature variation. These techniques can be applied at a row  and/or column, subarray, array, tile level.

\par




At weight level,  \textit{weight decomposition (WD)}, \textit{bit-width downgrading (BD)} and \textit{weight compensation (WC)} techniques are available.  These approaches are based on the ReRAM feature that implies that high conductance states are more vulnerable to an increase of temperature. To represent weights of different polarities, negative and positive crossbar arrays are used. The WD technique searches for the decomposition case so that the temperature distribution in both arrays is uniform and as low as possible. For further optimization, a \textit{fine-grained weight adjustment} method can be applied. BD is the dynamic thermal management method and is used when the temperature is above 330K. In this technique weight adjustment is performed by shifting bits and therefore can be applied only on weights with high resolution. Generally, BD and WC are the same operation; both methods shift conductance states and restore multiplication results, but WC is applied only on weights that were not protected by the WR and WPS techniques. 

Apart from manipulations with RCAs, \textit{Solution 8} suggested optimizing the layout by dis-centralizing the hot components like ADCs, DACs and eDRAM. Despite the seeming advantages, such implementation requires additional research since it leads to other challenges, e.g., reconsideration of routing and latency. Most importantly, the new layout may be incompatible with adopted chip fabrication standards. The temperature-aware training of the DNN model implies the resilience of the trained model to the ambient temperature change of a given range. In \textit{Solution 10}, the DNN model was trained with noise-injection. considering possible temperature fluctuations. Such a model remains resilient to temperature variations a certain period of time after mapping to the hardware and does not require retraining and reprogramming of the states\citet{meng2021temperature}.  One of the ways to improve the PKD method is to implement via injection lower noise levels to fully connected layers of CNN, as they are found to be more sensitive to noises \citet{yang2021multi}. 
\par 
Overall, the design of the majority of \textit{solutions} from Section \ref{sec:solutions} aimed to improve the accuracy and lifetime of ReRAM. Their fair comparison  on the same DNN model was not possible due to the unavailability of open-source codes. Some of the proposed solutions \citet{liu2019hr}, \citet{beigi2019thermal}, \citet{chen2022wrap}, \citet{abbey2022thermal} were designed for \textit{inference} of different workloads based on ISAAC configuration which uses \textit{naive and straightforward weight mapping} \citet{shafiee2016isaac}.  Later designs of accelerators \citet{qiao2018atomlayer},\citet{ankit2019puma} introduced weight \textit{reuse mapping} and support of \textit{training phase}  \citet{qiao2018atomlayer}, \citet{ankit2019puma}, \citet{song2017pipelayer} that highlights the need for reevaluation of the weight remapping methods.

Although the majority of the proposed optimization techniques were designed for HMC-like 3D configurations, it was noticed that remapping techniques did not take into account the impact of heat from neighboring "aggressor" cells in horizontal, vertical and diagonal directions. 
Moreover, in addition to information from thermal sensors, consideration of input distribution, the ReRAM cell's feature size, RCA proximity to ADC, DAC and eDRAM, and their pitch lengths would improve the weight reordering algorithms. Such a close relationship between temperature, ReRAM technology, architecture design and performance suggests that one of the best ways of developing thermal-aware and robust design should be solved as a MOO problem as in the case of AccuReD in \textit{Solution 9}. As mentioned earlier, AccuReD is a heterogeneous ReRAM/GPU platform that supports both inference and training. Unlike with the majority of other accelerators, the presence of full-precision GPU in AccuReD allows execution of \textit{Normalization} (V-norm) and \textit{SoftMax} layers and achieves near-GPU accuracy. Along with the pipeline latency and model accuracy, its weight mapping strategy takes into account vertical and horizontal heat flows as objectives. The authors also highlight that the MOO design and optimization problem can include other objectives and be solved by different MOO solvers.

Other methods for decreasing the effect of temperature on the hardware, and ReRAM-based designs in particular, include adding microfluidic cooling layers \citet{zhang20123d}. In \citet{peng2021heterogeneous} the authors proposed electrical-thermal co-design of a multitier CIM accelerator based on heterogeneous 3D integration (H3D) using TSV. Here, the number and diameter of TSVs were varied to find an optimal point between system performance and thermal disturbance. Besides, the number of tiers in the 3D structure was also considered a variable parameter. In \citet{joardar2020accured} TSV-based 3D design allows four tiers and M3D integration has up to eight tiers when threshold temperature is set to 373K, and therefore a preference is given to M3D due to faster heat dissipation. On top of that, one of the recent works \citet{sun2022multi} proposes benefiting from temperature and using natural biomaterials for manufacturing sustainable and pollution-free temperature-controlled ReRAM devices. These can be applied for the production of temperature-controlled sensors and detectors as well as medical treatment devices.

\bibliographystyle{unsrtnat}
\bibliography{ref}

\end{document}